\documentclass[twocolumn, journal]{IEEEtran}
\usepackage{graphicx}
\usepackage{amsmath,amssymb} % define this before the line numbering.
\usepackage{color}
\usepackage{times}
\usepackage{epsfig}
\usepackage{graphicx}
\usepackage{amsmath}
\usepackage{amssymb}
\usepackage{epstopdf}
\usepackage{algorithm}
\usepackage{algorithmic}
\usepackage{subfigure}
\usepackage{booktabs}
\usepackage{bm}
\usepackage[table]{xcolor}
\usepackage{hhline}
\usepackage{pgf}
\usepackage{tikz}
\usepackage{collcell}
\usepackage{multirow}
\usepackage{url}

\def\colorModel{hsb} %You can use rgb or hsb
\newcommand\ColCell[1]{
  \pgfmathparse{#1<50?1:0}  %Threshold for changing the font color into the cells
    \ifnum\pgfmathresult=0\relax\color{white}\fi
  \pgfmathsetmacro\compA{0}      %Component R or H
  \pgfmathsetmacro\compB{#1/100} %Component G or S
  \pgfmathsetmacro\compC{1}      %Component B or B
  \edef\x{\noexpand\centering\noexpand\cellcolor[\colorModel]{\compA,\compB,\compC}}\x #1
  } 
\newcolumntype{E}{>{\collectcell\ColCell}m{0.4cm}<{\endcollectcell}}  %Cell width
\newcommand*\rot{\rotatebox{90}}

\newcommand{\smallcolwidth}{11}
\ifCLASSINFOpdf
\else
\fi
\begin{document}
\title{Triple-view Convolutional Neural Networks for COVID-19 Diagnosis with Chest X-ray }

%\author{Michael~Shell,~\IEEEmembership{Member,~IEEE,}
%        John~Doe,~\IEEEmembership{Fellow,~OSA,}
%        and~Jane~Doe,~\IEEEmembership{Life~Fellow,~IEEE}% <-this % stops a space
%\thanks{M. Shell was with the Department
%of Electrical and Computer Engineering, Georgia Institute of Technology, Atlanta,

\author{Jianjia Zhang,
        Luping Zhou,~
        Lei Wang,~
        %Yang Wang,~\IEEEmembership{Senior Member,~IEEE,}
        %and Fang Chen,~\IEEEmembership{Senior Member,~IEEE,}
\thanks{Jianjia Zhang is with the School of Computer Science, University of Technology Sydney, Sydney, NSW 2007, Australia,
 e-mail: Jianjia.zhang@uts.edu.au.}%Yang.wang, Fang.chen <-this % stops a space
\thanks{Luping Zhou is with School of Electrical and Information Engineering, The University of Sydney, Sydney, NSW 2006, Australia,
e-mail:  luping.zhou@sydney.edu.au}% <-this % stops a space
\thanks{Lei Wang is with School of Computing and Information Technology, University of Wollongong, Wollongong, NSW 2522, Australia,
e-mail: leiw@uow.edu.au}% <-this % stops a space
}


\markboth{}%
{Shell \MakeLowercase{\textit{et al.}}: Bare Demo of IEEEtran.cls for IEEE Journals}



\maketitle

\begin{abstract}
%The abstract goes here.
The Coronavirus Disease 2019 (COVID-19) is affecting increasingly large number of people worldwide, posing significant stress to the health care systems. Early and accurate diagnosis of COVID-19 is critical in screening of infected patients and breaking the person-to-person transmission. Chest X-ray (CXR) based computer-aided diagnosis of COVID-19 using deep learning becomes a promising solution to this end. However, the diverse and various radiographic features of COVID-19 make it challenging, especially when considering each CXR scan typically only generates one single image. Data scarcity is another issue since collecting large-scale medical CXR data set could be difficult at present. Therefore, how to extract more informative and relevant features from the limited samples available becomes essential. To address these issues, unlike traditional methods processing each CXR image from a single view, this paper proposes triple-view convolutional neural networks for COVID-19 diagnosis with CXR images. Specifically, the proposed networks extract individual features from three views of each CXR image, i.e., the left lung view, the right lung view and the overall view, in three streams and then integrate them for joint diagnosis. The proposed network structure respects the anatomical structure of human lungs and is well aligned with clinical diagnosis of COVID-19 in practice. In addition, the labeling of the views does not require experts' domain knowledge, which is needed by many existing methods. The experimental results show that the proposed method achieves state-of-the-art performance, especially in the more challenging three class classification task, and admits wide generality and high flexibility.

\end{abstract}


% \fi
\IEEEpeerreviewmaketitle

\section{Introduction}
%\IEEEPARstart{T}{his} demo file is intended to serve as a ``starter file''
%for IEEE journal papers produced under \LaTeX\ using
%IEEEtran.cls version 1.8b and later.
%I wish you the best of success.
\IEEEPARstart{T}{he} Coronavirus Disease 2019 (COVID-19), caused by severe acute respiratory syndrome coronavirus 2 (SARSCoV-2), is quickly spreading over the world and a huge number of people have been affected.  Nearly 27 million COVID-19 cases and 0.9 million deaths have been confirmed globally as of 07 Sep. 2020~\cite{world2020coronavirus}. And the numbers still keep rising dramatically due to high rate of community transmission and lack of appropriate treatment and vaccines~\cite{oh2020deep}. A critical measure to stop the virus transmission and restrain the outbreak is early diagnosis~\cite{ouyang2020dual,mei2020artificial}. Early diagnosis not only enables timely treatment for the patients affected, but also allows quick isolation of their close contacts for disease containment~\cite{mei2020artificial}.

Although real-time reverse-transcriptionpolymerase-chainreaction (RT-PCR) is considered as the golden standard to make a definitive diagnosis of COVID-19 infection, it encounters several issues in such a global pandemic.  Firstly, its false negative rate is high. A patient initially tested negatively could be later tested positive~\cite{zu2020coronavirus,chan2020familial}. Therefore, a series of RT-PCR tests, which can take up to two days~\cite{mei2020artificial}, may be required to confirm a case. Secondly, RT-PCR test kits may not be sufficiently available in all areas across the world, especially when considering that the global supply chains are at risk due to major disruptions. In addition, the laboratory equipment required by RT-PCR could be also a bottleneck to conduct large scale tests. These issues may result in delayed or even missed diagnosis and a computer-aided diagnosis system could, at least partially, automate the diagnosis and facilitate large scale screening of COVID-19 patients. %a natural question arises: is it possible to develop a computer-aided diagnosis system to, at least partially, automate the diagnosis and facilitate large scale screening of COVID-19 patients?

%An investigation on the effects of COVID-19 on patients could help to answer the question. 
SARSCoV-2 infection can attack various types of  lung cells~\cite{li2020sars} and trigger an inflammatory response in the air sacs of lungs, leading to a typical symptom of COVID-19 patients: pneumonia. Making matters even worse, both the left and right lungs are often involved for early, intermediate and late stage patients~\cite{9086482}, causing breathlessness and even death. The inflammatory response can be detected by radiology examinations, especially chest computed tomography (CT) or chest X-ray (CXR). Typical radiographic features of COVID-19 patients in these scans include ground-glass opacities (GGO), multifocal patchy consolidation and/or interstitial changes with a peripheral distribution~\cite{ouyang2020dual,chung2020ct}. %as shown in the example CXR images from COVID-19 patients in Fig.~\ref{fig:xrayexample} in comparison with normal subjects and Other lung disease patients.
These visual features specific to COVID-19 patients are used by clinicians for COVID-19 diagnosis. At the same time, they admit the possibility of computer-aided diagnosis. There have been considerable research interests devoted to it and the existing works can be summarized as follows from four different perspectives: 1) \textbf{Radiology exam used}: CXR~\cite{oh2020deep,basu2020deep,pereira2020covid,luz2020towards,wang2020covid,horry2020x} and CT~\cite{9086482,ouyang2020dual,bai2020ai} scans are the most commonly used radiology exams by recent works for computer-aided diagnosis of COVID-19. Ultrasound is also explored in a few recent studies, such as~\cite{roy2020deep,soldati2020there}. At the same time, some other studies explore a combination of multiple source data for joint learning. For example, the works in~\cite{mei2020artificial,zhang2020clinically} integrate CT scans with non-imaging clinical metadata, e.g., clinical symptoms, exposure history, and/or laboratory testing, for COVID-19 prediction. 
2) \textbf{Diagnostic objectives}: The existing works can be categorized into three groups according to their aims. The first group, e.g., in~\cite{pereira2020covid,oh2020deep,luz2020towards,wang2020covid,horry2020x,zhang2020clinically}, is to differentiate normal controls, COVID-19 patients, and other non-COVID-19 pneumonia, e.g., Severe Acute Respiratory Syndrome (SARS) and Middle East Respiratory Syndrome (MERS), which forms a three class classification problem. The second group~\cite{9086482,ouyang2020dual,bai2020ai} aims to solve a binary classification problem of distinguishing  COVID-19 from other non-COVID-19 pneumonia. The remaining group~\cite{mei2020artificial} also works on a binary classification problem, but it focuses on recognizing COVID-19 from normal. 
3) \textbf{Labeling information}: Some existing works, e.g., \cite{pereira2020covid,9086482,luz2020towards,wang2020covid} only use the ground truth diagnosis results as labels in the training stage. Meanwhile, many other works~\cite{oh2020deep,ouyang2020dual,mei2020artificial,horry2020x,zhang2020clinically} also utilize the lung mask or lesion segmentation results to train a more focused classifier. For example, the work in~\cite{ouyang2020dual} proposes an online attention module in Convolutional neural networks (CNN) to focus on the infection regions in lungs when making diagnosis decisions. 
4) \textbf{Shallow or deep models}: Most of existing works~\cite{roy2020deep,oh2020deep,ouyang2020dual} use CNN in developing COVID-19 diagnosis model. Shallow models are also explored in a few works~\cite{pereira2020covid,9086482} with hand-crafted features or combined with deep models as in~\cite{mei2020artificial,zhang2020clinically}.

There are several challenging issue faced by the earlier studies. One critical issue is that the radiographic features of COVID-19 are diverse and can vary \cite{shi2020radiological}. Another challenge is the scarcity of training data since collecting large scale training data is difficult, if not impossible, in the current emergent situation of COVID-19 pandemic. Moreover, manual segmentation of lung or lesion masks for each scan required by the works in \cite{oh2020deep,ouyang2020dual,mei2020artificial,horry2020x,zhang2020clinically} not only is costly and time-consuming, but also requires experts' domain knowledge. These issues can affect the scalability and generality of the models. In this case, it is desirable to design an effective model using limited training data without requirement of domain knowledge. This is attempted this paper.

%The primary motivation of TV-CovNet is to extract richer and more discriminative features from a limited training set available to improve the diagnosis accuracy. 

Inspired by the encouraging success achieved by the earlier studies and motivated by the issues we are facing, this work proposes a novel triple-view CNN for COVID-19 diagnosis with CXR images and we denote it as TV-CovNet in the remaining sections.   Specifically, the proposed TV-CovNet consists of three streams with three views of the lungs in each CXR image, i.e., the left lung view, the right lung view and the overall view, as inputs and conducts diagnosis by integrating the information from them. This idea respects the anatomical lung structure and pathology of COVID-19. The bronchi of the left and right lungs are internally connected by the trachea, therefore the SARSCoV-2 could easily transit form one lung to the other, especially when considering the high transmission rate of COVID-19. That's why bilateral lung involvement can often be observed for COVID-19 patients. The proposed TV-CovNet is also inspired by clinical practice in diagnosing COVID-19 with CXR images, i.e., a clinician usually checks the left and right lungs individually and then jointly before making a decision. In comparison with traditional methods which consider each CXR image as an single-view image, the triple-view structure and joint decision making enable TV-CovNet to extract more representative and relevant features from each CXR image. This is especially important when the training data are limited. Moreover, for each CXR image at either the training or the test stage, we only need to provide three lung bounding boxes and no expert knowledge from clinicians is required. This improves the generality of the proposed method in comparison with the methods in~\cite{oh2020deep,ouyang2020dual,mei2020artificial,horry2020x,zhang2020clinically}, which highly rely on accurate contour segmentation of  lung or lesion areas. As far as we are aware, we are among the first ones to investigate the individual left and right lung views for CXR based COVID-19 diagnosis. CXR scan is used in the proposed TV-CovNet since it is cheaper, faster and more accessible across the world in comparison with CT.

This paper's main contributions are summarized as follows:
\begin{itemize}
\item We propose a novel triple-view CNN to conduct COVID-19 diagnosis using CXR images. The proposed networks can extract and integrate the diagnostic clues from the left lung, right lung and overall lung views for a joint decision, which respects the anatomical structure of lungs and is well aligned with the practical diagnosis by clinicians.
\item Various methods are explored to integrate the information from the three views in a fusion layer. As will be demonstrated in the experimental evaluation, the average pooling at score layer attains the best performance.
\item The superior performance of the proposed method over the competing methods and its wide generality are consistently verified in two tasks with three backbone network architectures. Specifically, one task is differentiating normal from COVID-19, and the other is distinguishing normal, COVID-19,  and other non-COVID-19 pneumonia. Most recent works only study one of these two tasks, but both of  them are covered in this paper since we believe that either one may be important in certain applications. The three backbone network architectures refer to ResNet-50, ResNet-101 and ResNet-152. %, which are utilized in TV-CovNet respectively.

\item We also demonstrate the high flexibility of the proposed network, i.e., it could be easily integrated with other state-of-the-art deep learning methods to further improve the diagnosis performance.
\end{itemize}

\section{Related Works}
Motivated by their success in computer vision tasks, e.g., object detection and image classification, deep learning has been intensively studied for  diagnosis of various conditions ranging from breast lesions~\cite{cheng2016computer}, cardiac disease~\cite{bernard2018deep} to, the focus of this paper, pneumonia~\cite{rajpurkar2017chexnet} in recent years. Due to the emergent COVID-19 pandemic at present, there is intensive research interest devoting to computer-aided COVID-19 diagnosis using deep learning with radiology imaging. 

As one of the main complications caused by COVID-19, pneumonia is an infection of the lung tissues, including the bronchi, bronchioles and alveoli, resulting in breathing difficulties or even respiratory failure. CXR~\cite{oh2020deep,basu2020deep,pereira2020covid,luz2020towards,wang2020covid,horry2020x} and CT~\cite{9086482,ouyang2020dual,bai2020ai,mei2020artificial,zhang2020clinically} scans are the most commonly used radiologic examinations by clinicians in identification of pneumonia inflammation. COVID-19 positive cases  present radiographic abnormalities such as ground-glass opacity and bilateral patchy shadowing in CXR and CT images~\cite{guan2020clinical,wang2020covid}. 
Although CT scan could provide more detailed diagnostic clues in identification of COVID-19 positive cases, CXR is probably a more practical option, especially in resource-constrained or heavily-affected areas for its various advantages over CT~\cite{wang2020covid}. Firstly, X-ray machines are more widely and readily accessible across the world. It is one of the most equipped device in all levels of medical institutions due to its wide application and significantly lower price in comparison with CT. 
%since it is much cheaper than CT equipment and is a standard equipment . %One CT equipment can cost more than ten times of a X-ray equipment~\cite{pereira2020covid}. 
Secondly, CXR has higher scanning efficiency, admitting rapid screening. Thirdly, there are various kinds of portable CXR machines, which can better adapt to various application contexts. Last but not least, a CXR scan delivers much less dose of radiation in comparison with a CT scan, typically less than 4\% of the later~\cite{kim2016comparison}. These advantages motivate this paper to develop a CXR based COVID-19 diagnosis method.

While diagnosis via CXR scan enjoys various advantages as explained above, it also brings challenges to develop a robust and effective diagnosis method. On the one hand, each CXR scan typically only generates one single image for diagnosis. In comparison, a CT scan generates 3D volumetric data and enable detailed visualization from multiple perspectives. On the other hand, the CXR scans from COVID-19 patients that can be used as training data are limited since it is difficult to collect large scale of samples considering the current emergent situation. In this case, how to better extract informative features from each sample becomes essential. Many existing works~\cite{oh2020deep,horry2020x} resort to manual labeling of lesion or lung masks, however, this is time-consuming and can only be done by experts. This paper attempts to address this issue without requirement of experts in labeling from another perspective. That is, unlike the existing works which take a CXR scan as a single view image, to explore multi-view feature extraction from each CXR scan by building multi-stream networks. 

In fact, there has been a number of attempts to develop multi-stream networks in computer vision tasks, e.g., action recognition~\cite{simonyan2014two}  %,feichtenhofer2016convolutional,zolfaghari2017chained 
and 3D reconstruction~\cite{gundogdu2019garnet}. In these tasks, the input data can be naturally decomposed into multiple components. For example, a video clip of action sequence is split into  spatial and temporal components in~\cite{simonyan2014two}. In the area of medical data analysis, a similar methodology has been explored  to fuse multi-stream information within shallow models~\cite{liu2015inherent,zhang2017multi,9086482} or deep learning models~\cite{jin2019accurate,mei2020artificial}. These works can be roughly categorized into two groups: i) the first group aims to fuse different modalities, e.g., MRI and PET data in~\cite{zhang2017multi}, PET and CT in~\cite{jin2019accurate} and CT and clinical meta data in~\cite{mei2020artificial}; ii) the second group works on combining multiple features extracted from the same modality, e.g., features with multiple templates of brain regions of interest in~\cite{liu2015inherent} and multiple hand-crafted features from CT images in~\cite{9086482}. These works consistently demonstrate that a fusion of multiple modalities or features is able to improve the performance over any single modality or feature. Unfortunately, these multi-stream networks can hardly be readily applied to sole CXR scan, upon which we are trying to develop computer-aided diagnosis method in this paper. One obvious reason is that we presume no other modality scan or clinical information is available except a single CXR image per subject. In addition, although a combination of multiple hand-crafted features in shallow models is always doable, CNN based end-to-end methods are more preferable due to their capability to learn more specific and adaptive features, especially for challenging CXR images. 

After a careful review of the existing works and the challenges as explained above in this line, we propose TV-CovNet for COVID-19 diagnosis with CXR images. %The triple-view means the left lung view, the right lung view and the overall view. 
As aforementioned, the proposed TV-CovNet is in alignment with both of the anatomical lung structures and clinicians' practical diagnosis of COVID-19. %The human lungs are naturally composed of left and right lungs, which are internally connected by the trachea. Therefore, the SARSCoV-2 could easily transit from one side to the other, which explains why bilateral lung involvement can be observed in most COVID-19 patients. 

\section{Method}
\begin{figure*}[!htb]
\begin{center}
{\includegraphics[width = 160 mm]{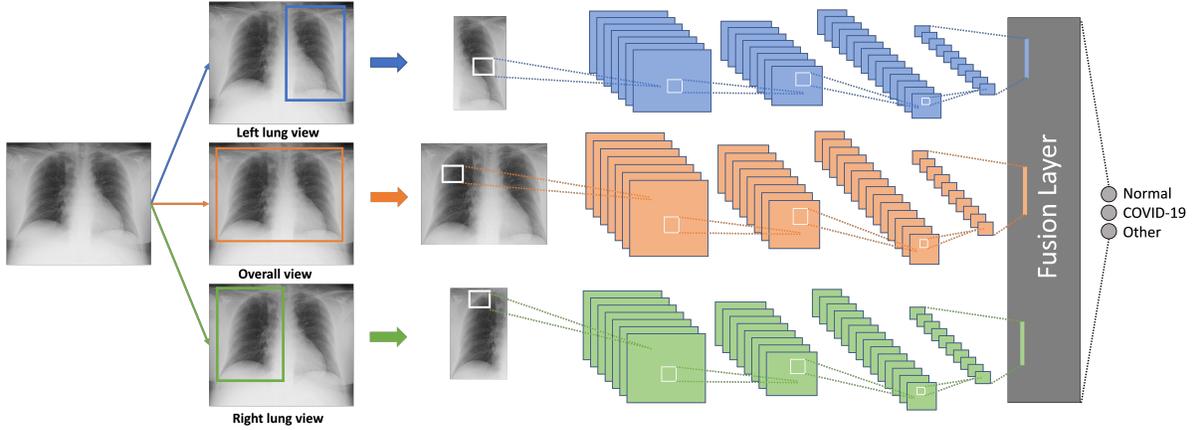}}
\end{center}
\vspace{-6 mm}
\caption{Overview of the proposed triple-view Convolutional Neural Networks for COVID-19 Diagnosis with Chest-X-ray.}
\label{fig:flowchart}
\end{figure*}

The overall framework of the proposed TV-CovNet is illustrated in Fig~\ref{fig:flowchart}. As seen, each input CXR image, which is assumed in  posteroanterior (PA) view in this paper, is firstly cropped into the left lung view (the top blue stream), the overall view (the middle orange stream) and the right lung view (the bottom green stream). In comparison with lesion or lung mask labeling in the literature, such as in \cite{oh2020deep,ouyang2020dual,mei2020artificial,horry2020x,zhang2020clinically}, this cropping step is efficient and no domain expert knowledge is required. Therefore, this cropping step is not restricted to clinicians and this could significantly improve the generality of the proposed method. Also, it will still be applicable when large scale of training data is available in future. Once cropped, the three views are fed into the three streams of TV-CovNet respectively. As shown in Fig~\ref{fig:flowchart}, three colors, i.e., blue, orange and green, are used to indicates the corresponding streams. The backbone network architecture of each stream is flexible, i.e., be it a  specifically-designed or off-the-shelf network architecture. And the network architectures in the three streams can be identical or different. The features extracted from the three streams will be combined in a fusion layer, which will be explained in detail in the following section.  The fusion layer will be followed by a final fully connected classification layer. The advantages of the proposed TV-CovNet are recapped as follows: 

\begin{itemize}
\item In comparison with the existing works taking a single overall view of CXR scans as input, TV-CovNet could extract more detailed and complementary information from the left lung, right lung and the overall views in three different streams, providing more clues for accurate diagnosis. This is motivated by the clinical practice in diagnosing COVID-19. Each CXR scan of human lungs in PA view composes of the left and right lungs, and these two lungs may present different visual characteristics although they are internally connected and, in most COVID-19 cases, bilaterally affected. Therefore, the left and right lungs are usually reviewed individually and jointly by a clinician to identify more diagnostic clues;
\item TV-CovNet does not require experts in manual labeling of data, admitting high generality to different sized data sets. More detailed labeling information, e.g., lesion or lung mask labeling in \cite{oh2020deep,ouyang2020dual,mei2020artificial,horry2020x,zhang2020clinically}, could certainly help to train a more focused network and may improve the diagnosis accuracy. At the same time, however, it may limit the generality of the networks due to extensive requirement on experts' domain knowledge. In contrast, cropping three views in TV-CovNet is much simpler and more efficient without requirement of  domain knowledge;
\end{itemize}

In addition, as a byproduct, TV-CovNet could possibly alleviate the effects of dataset bias. In order to construct a CXR training dataset for CNN based COVID-19 diagnosis, most works in the literature combine multiple datasets from various institutes as one, as in \cite{oh2020deep,ouyang2020dual,pereira2020covid,luz2020towards}.   However, as pointed out in \cite{maguolo2020critic,pereira2020covid}, joining different databases may add bias from the non-lung areas,  which can be learned by the networks to recognize which origin database the test sample is from rather than the lung injuries, especially when the training data is scarce. To a certain extent, the proposed method could minimize the negative effects of this issue by only feeding the cropped lung areas in the left and right lung views. On the one hand, the left and right lung views are non-overlapping, so no specific non-lung area will appear in both of these two views. On the other hand, the strategy of diagnosing from triple views of the lungs will help to alleviate the effects of the non-lung areas in the overall view.
%in the overall view will not appear in both of these two views and the dataset bias introduced by them can be alleviated.

\begin{figure*}[!htb]
\begin{center}
{\includegraphics[width = 110 mm]{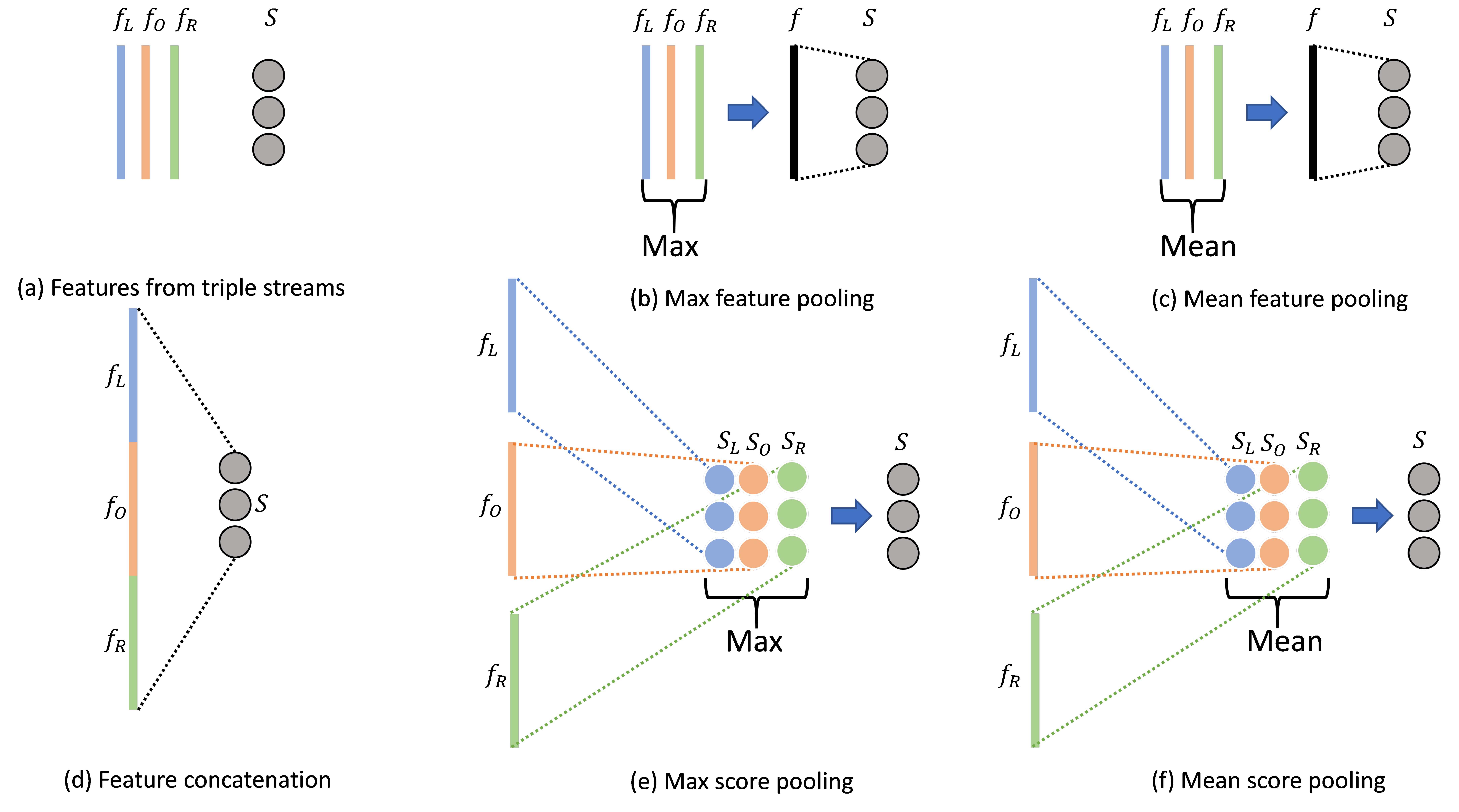}}
\end{center}
\vspace{-6 mm}
\caption{Illustration of the fusion methods in the proposed triple-view CNN  structure.}
\label{fig:pooling}
\end{figure*}

\subsection{Fusion methods} \label{sectionfusion}
The above section explains how TV-CovNet extracts features from three streams. With these output features, a following issue is how to effectively integrate them in a fusion layer, as illustrated in Fig.~\ref{fig:flowchart}. To address this issue, five fusion methods will be studied. Specifically, the five fusion methods performs at two levels, i.e., feature level (sub-figure (b), (c), (d) in Fig.~\ref{fig:pooling}) and score level ( sub-figure (e), (f) in Fig.~\ref{fig:pooling}). 

\textbf{Feature level fusion}. Let us denote the output feature vectors from the left lung, overall and right lung streams as $\bm{f}_{L}$, $\bm{f}_{O}$ and $\bm{f}_{R}$, respectively, and they are assumed to have the same dimensionality $D$, i.e., $\bm{f}_{L}$, $\bm{f}_{O}$, $\bm{f}_{R}$ $\in R^D$, as demonstrated in Fig.~\ref{fig:pooling} (a). This can be easily met by using the same backbone network architecture in the three streams or appending a fully connected layer with D-dimensional output if different network architectures are used in the three streams.

At this level, a fusion function $\mathcal{F}$: $\bm{f}_{L}$, $\bm{f}_{O}$, $\bm{f}_{R}$ $\to$ $\bm{f}$ combines the three features $\bm{f}_{L}$, $\bm{f}_{O}$, $\bm{f}_{R}$ into a combined feature $\bm{f}$. The combined feature $\bm{f}$ will be fed into the classification layer to calculate the class score $\bm{s}$. We investigate the following three fusion functions illustrated in Fig.~\ref{fig:pooling}. (b), (c), (d):
\begin{itemize}
\item \textbf{Max feature pooling.} $\bm{f} = \mathcal{F}_{max} (\bm{f}_{L}, \bm{f}_{O}, \bm{f}_{R}$) takes the element-wise maximum of the three input feature vectors:\\
$\bm{f}_{d} = max\{\bm{f}_{L,d}, \bm{f}_{O,d}, \bm{f}_{R,d}\}$, \\
where $d \in [1,D]$ indicates the $d$-th feature dimension. The resulting feature $\bm{f}$ after max pooling operation is still a vector with the same dimension $D$ as input vectors.
\item \textbf{Mean feature pooling.} $\bm{f} = \mathcal{F}_{mean} (\bm{f}_{L}, \bm{f}_{O}, \bm{f}_{R}$) takes the mean of the three input feature vectors:\\
$\bm{f}_{d} = mean\{\bm{f}_{L,d}, \bm{f}_{O,d}, \bm{f}_{R,d}\}$, with $d \in [1,D]$\\
Similarly, the resulting feature $\bm{f}$ after mean pooling is also a $D$-dimensional vector.
\item \textbf{Feature concatenation.} $\bm{f} = \mathcal{F}_{cat} (\bm{f}_{L}, \bm{f}_{O}, \bm{f}_{R}$) stacks the three input feature vectors:\\
$\bm{f}_{d} = {\bm{f}_{L,d}}$ when $d \in [1,D]$\\
$\bm{f}_{d} = {\bm{f}_{O,d-D}}$ when $d \in [D+1,2D]$\\
$\bm{f}_{d} = {\bm{f}_{R,d-2D}}$ when $d \in [2D+1,3D]$\\
The resulting concatenated feature $\bm{f}$ is $3D$-dimensional. 
\end{itemize}
\textbf{Score level fusion}. 
If fusion is applied at the score level, the scores of the last classification layer in the three streams will be combined. Let $\bm{s}_{L}$, $\bm{s}_{O}$ and $\bm{s}_{R}$ denote the corresponding class score vectors from the left, overall and right lung streams.

At this level, a fusion function  $\mathcal{F}$: $\bm{s}_{L}$, $\bm{s}_{O}$, $\bm{s}_{R}$ $\to$ $\bm{s}$ combines the class score vectors  $\bm{s}_{L}$, $\bm{s}_{O}$, $\bm{s}_{R}$ into the final class score $\bm{s}$. %The combined score $\bm{s}$ will determine the classification results. 
We investigate the following two fusion functions, as illustrated in sub-figures (e), (f) in Fig.~\ref{fig:pooling}:
\begin{itemize}
\item \textbf{Max score pooling.} Similar to the \textbf{Max feature pooling} above, $\bm{s} = \mathcal{F}_{max} (\bm{s}_{L}, \bm{s}_{O}, \bm{s}_{R}$) takes the element-wise maximum of the three input score vectors:\\
$\bm{s}_{i} = max\{\bm{s}_{L,i}, \bm{s}_{O,i}, \bm{s}_{R,i}\}$, \\
where $i$ indicates the $i$-th class.
\item \textbf{Mean score pooling.} Similarly, $\bm{s} = \mathcal{F}_{mean} (\bm{s}_{L}, \bm{s}_{O}, \bm{s}_{R}$) takes the mean of the three input score vectors:\\
$\bm{s}_{i} = mean\{\bm{s}_{L,i}, \bm{s}_{O,i}, \bm{s}_{R,i}\}$. 
\end{itemize}

\section{Evaluation}
\subsection{Dataset}\label{dataset}
The dataset used in our evaluation is constructed by CXR images from the following two publicly accessible sources:
\begin{itemize}
\item \textbf{COVID-19 image data collection} \cite{cohen2020covid}. This dataset contains CXR images of patients which are confirmed as COVID-19 or other non-COVID-19 pneumonia. All the CXR images in PA view are collected into the combined dataset. The COVID-19 cases form the ``COVID-19'' class while the other non-COVID-19 pneumonia cases are assigned to the ``Other''  class.
\item \textbf{Chest X-ray Database} \cite{candemir2013lung,jaeger2013automatic}. This CXR dataset is composed of normal cases and patients with tuberculosis. %This data created by the National Library of Medicine, Maryland, USA in collaboration with Shenzhen No.3 People’s Hospital, Guangdong Medical College, Shenzhen, China. 
The normal cases form the ``Normal'' class and tuberculosis cases are assigned to the ``Other'' class.
\end{itemize}
The numbers of CXR samples in each class of the combined dataset are shown in Table~\ref{tab:datasetnumber}. The left lung, right lung and overall views of each CXR scan, as shown in Fig.~\ref{fig:flowchart}, in the combined dataset will be cropped.
\begin{table}[!ht]
\caption{Number of CXR scans in each class of the combined dataset.}
\label{tab:datasetnumber} \centering
\begin{tabular}{l|c|c|c|c}
class&Normal&COVID-19&Other&\textbf{Total}\\
\hline
number&326&217&389&\textbf{932}\\
%Tuberculosis               &336\\
%SARS              &16\\
%Streptococcus     &13\\
%Pneumocystis      &12\\
%E.Coli             &4\\
%ARDS              & 4\\
%Legionella         &2\\
%Klebsiella         &1\\
%Chlamydophila      &1\\
\hline
\end{tabular}
\end{table}

\subsection{Implementation Details}
We use  ResNet~\cite{he2016deep} architecture pretrained on ImageNet \cite{imagenet_cvpr09} as the backbone network due to its promising performance in recent works on COVID-19 diagnosis~\cite{luz2020towards,wang2020covid,horry2020x}. In order to verify the generality of the proposed TV-CovNet, three backbone networks are adopted in two classification tasks. Specifically, ResNet-50, ResNet-101 and ResNet-152 will be applied respectively within TV-CovNet to evaluate their performance. And the two classification tasks refer to a binary classification task between Normal and COVID-19 and a three class classification task among Normal, COVID-19 and Other. Cross entropy is used as the objective function with learning rate of 0.01 and momentum of 0.9.  The training process is iterated for 100 epochs with a batch size of 10 samples and the learning rate is decreased by a factor of 0.1 every 20 epochs. Among all the samples in each class, we randomly select 60\% of them as training set  and the rest as test set. This random training/test partition is repeated 15 times for each classification task to obtain stable statistics.

\subsection{Competing Methods} The following two recent works achieving state-of-the-art performance on COVID-19 diagnosis with CXR images will be involved in our experimental evaluation:
\begin{itemize}
\item \textbf{COVID-Net}~\cite{wang2020covid}. COVID-Net is one of the earliest  work on  designing CNN for the detection of COVID-19 cases with CXR images. It works on classifying normal, COVID-19 and non-COVID-19 pneumonia, which shares the same setting as that of the three class classification task in this paper. Their publicly accessible implementation on \url{https://github.com/lindawangg/COVID-Net} is used in our evaluation and the overall view of CXR images is used as input to train the networks.
\item \textbf{LocalPatch}~\cite{oh2020deep}. LocalPatch is another recent work closely related to ours. LocalPatch tries to address the issue of data scarcity by developing a patch-based CNN with many random patches cropped from each CXR image. The final classification result for a test sample is obtained by majority voting from inference results of the patchy crops from the test CXR image. We implemented their network by following the paper. As in the paper, 100 random patches are cropped from the overall view of each CXR scan for majority voting during the inference stage. %Lung masks are not used in this method for fair comparison.
\end{itemize}

\subsection{Preliminary Study}
The essential objective of this paper is to extract informative visual features from the left and right lung views in addition to the overall view to boost the diagnosis accuracy. From this sense, an implicit assumption behind is that the left or right lung view should contain certain information for diagnosis. In order to verify the validity of this assumption, this section explores COVID-19 diagnosis using a single stream networks with a single view of CXR images as input. Specifically, a pretrained ResNet, i.e., one of ResNet-50, ResNet-101 or ResNet-152,  is used as backbone networks and the final classification layer is reset to predict three classes of ``Normal'', ``COVID-19'' and ``Other''. Note that each classifier has only one stream as in the commonly used setting from the literature, and it is fine-tuned with samples from one of the overall, left lung or right lung views. The performance of the fine-tuned networks is reported in Table~\ref{tab:prestudy}. As seen, with the overall view, ResNet-50 achieves an accuracy of 79.4\%, which is reasonably good considering the challenges of CXR based COVID-19 diagnosis. If left or right lung view is used, ResNet-50 obtains 80.2\% and 79.6\%, respectively, which are comparable or even slightly higher than that of the overall view. This demonstrates that a single lung view indeed contains substantial valuable information for diagnosis. Moreover, it is interesting to see that the left or right lung view could even outperform the overall view considering that the left or right lung view is essentially an interior part of the overall view. This is probably because although the overall view contains all the visual contents in both of the left and right lung views, it also contains considerable non-lung areas, whose visual characteristics could introduce bias or disturbance to the classifier. In contrast, these non-lung areas are partially excluded from the left or right lung views, so the networks could focus more on the lung area and extract COVID-19 related features. Similar results can be observed when ResNet-101 or ResNet-152 is used. These results lay a solid foundation for integration of multiple views to extract complementary diagnostic information and alleviate the effects of non-related features. 

\begin{table}[!ht]
\caption{Classification accuracy (\%) on three class COVID-19 diagnosis with unique view.}
\label{tab:prestudy} \centering
\begin{tabular}{l|c|c|c}
Network&Overall view&Left lung view&Right lung view\\
\hline
ResNet-50&79.4&80.2&79.6\\
ResNet-101&78.8&79.1&80.5\\
ResNet-152&78.9&79.1&78.9\\
\end{tabular}
\end{table}

\begin{table*}[!ht]
\caption{Comparison on COVID-19 Diagnosis With Chest X-ray Imaging (Two experiments).}
\label{tab:covidsummary} \centering
\begin{tabular}{l|ccccc|c}
					 & \multicolumn{5}{c}{$2$ classes} & $3$ classes \\
Methods in comparison & Accuracy & Precision& Recall &Specificity  &F1 score & Accuracy \\
\hline
COVID-Net~\cite{wang2020covid}&-&-&-&-&-&$78.0$ $\pm 2.1$\\
{ResNet-18 + LocalPatch}~\cite{oh2020deep}&$99.4$ $\pm 0.5$&$99.2$ $\pm 1.0$&$99.4$ $\pm 0.8$&$99.4$ $\pm 0.7$&$99.3$ $\pm 0.7$& $80.5$ $\pm 1.9$\\
\hline
\hline
{ResNet-50}~\cite{he2016deep}&$98.7$ $\pm 0.8$&$98.3$ $\pm 1.8$&$98.5$ $\pm 1.5$&$98.9$ $\pm 1.2$&$98.4$ $\pm 1.0$& $79.4$ $\pm 2.2$\\
{ResNet-50 + LocalPatch}~\cite{oh2020deep}&$99.7$ $\pm 0.4$&\bm{$99.7$} $\pm 0.5$&$99.6$ $\pm 0.7$&\bm{$99.8$} $\pm 0.4$&$99.6$ $\pm 0.5$& $81.6$ $\pm 1.8$\\
%\hline
TV-CovNet\_50\_fea\_cat& $97.9$ $\pm 1.2$&$97.1$ $\pm 1.9$&$97.7$ $\pm 1.8$&$98.1$ $\pm 1.3$&$97.4$ $\pm 1.4$& $79.3$ $\pm 2.0$\\
TV-CovNet\_50\_fea\_max& $97.8$ $\pm 1.2$&$98.0$ $\pm 2.0$&$96.6$ $\pm 2.1$&$98.7$ $\pm 1.2$&$97.2$ $\pm 1.6$& $80.4$ $\pm 2.2$\\
TV-CovNet\_50\_fea\_mean& $99.1$ $\pm 0.6$&$99.0$ $\pm 1.1$&$98.7$ $\pm 1.0$&$99.3$ $\pm 0.8$&$98.9$ $\pm 0.8$& $82.1$ $\pm 2.4$\\
TV-CovNet\_50\_sc\_max& $98.3$ $\pm 1.2$&$98.4$ $\pm 1.3$&$97.5$ $\pm 2.4$&$98.9$ $\pm 0.8$&$97.9$ $\pm 1.6$& $79.0$ $\pm 2.6$\\
TV-CovNet\_50\_sc\_mean& \bm{$99.8$} $\pm 0.3$&{$99.5$} $\pm 0.6$&\bm{$99.9$} $\pm 0.3$&{$99.7$} $\pm 0.4$&\bm{$99.7$} $\pm 0.4$& \bm{$83.7$} $\pm 2.0$\\
\hline
\hline
{ResNet-101}~\cite{he2016deep}&$98.7$ $\pm 1.1$&$98.2$ $\pm 2.2$&$98.6$ $\pm 1.2$&$98.7$ $\pm 1.7$&$98.4$ $\pm 1.2$& $78.8$ $\pm 2.5$\\
{ResNet-101 + LocalPatch}~\cite{oh2020deep}&\bm{$99.7$} $\pm 0.3$&$99.6$ $\pm 0.6$&\bm{$99.8$} $\pm 0.5$&$99.7$ $\pm 0.4$&\bm{$99.7$} $\pm 0.4$& $81.1$ $\pm 1.7$\\           
%\hline
TV-CovNet\_101\_fea\_cat& $97.8$ $\pm 1.1$&$97.3$ $\pm 1.9$&$97.1$ $\pm 2.3$&$98.2$ $\pm 1.3$&$97.2$ $\pm 1.4$& $78.8$ $\pm 2.8$\\
TV-CovNet\_101\_fea\_max& $97.8$ $\pm 1.3$&$97.0$ $\pm 1.9$&$97.6$ $\pm 2.4$&$97.9$ $\pm 1.3$&$97.3$ $\pm 1.7$& $80.5$ $\pm 2.2$\\
TV-CovNet\_101\_fea\_mean& $99.0$ $\pm 0.6$&$98.6$ $\pm 1.5$&$98.9$ $\pm 1.1$&$99.1$ $\pm 1.0$&$98.7$ $\pm 0.7$& $81.2$ $\pm 1.9$\\
TV-CovNet\_101\_sc\_max& $98.1$ $\pm 1.3$&$98.2$ $\pm 1.2$&$96.9$ $\pm 2.6$&$98.8$ $\pm 0.8$&$97.5$ $\pm 1.8$& $ 79.4$ $\pm 2.2$\\
TV-CovNet\_101\_sc\_mean& {$99.6$} $\pm 0.4$&\bm{$99.7$} $\pm 0.5$&{$99.4$} $\pm 0.9$&\bm{$99.8$} $\pm 0.4$&{$99.6$} $\pm 0.5$& \bm{$83.0$} $\pm 2.0$\\
\hline
\hline
{ResNet-152}~\cite{he2016deep}&$98.4$ $\pm 1.2$&$98.3$ $\pm 1.4$&$97.8$ $\pm 2.3$&$98.8$ $\pm 1.0$&$98.0$ $\pm 1.5$& $78.8$ $\pm 1.9$\\
{ResNet-152 + LocalPatch}~\cite{oh2020deep}&\bm{$99.8$} $\pm 0.3$&\bm{$99.6$} $\pm 0.5$&\bm{$99.9$} $\pm 0.3$&\bm{$99.7$} $\pm 0.4$&\bm{$99.8$} $\pm 0.4$& $81.4$ $\pm 2.1$\\
%\hline
TV-CovNet\_152\_fea\_cat& $98.4$ $\pm 0.9$&$97.9$ $\pm 1.8$&$98.2$ $\pm 1.5$&$98.6$ $\pm 1.2$&$98.0$ $\pm 1.0$& $79.1$ $\pm 2.1$\\
TV-CovNet\_152\_fea\_max& $97.8$ $\pm 1.7$&$97.6$ $\pm 2.2$&$97.0$ $\pm 3.3$&$98.4$ $\pm 1.5$&$97.3$ $\pm 2.2$& $80.4$ $\pm 3.3$\\
TV-CovNet\_152\_fea\_mean& $98.9$ $\pm 1.2$&$98.8$ $\pm 1.6$&$98.5$ $\pm 2.6$&$99.2$ $\pm 1.1$&$98.6$ $\pm 1.6$& $81.3$ $\pm 2.0$\\
TV-CovNet\_152\_sc\_max& $98.3$ $\pm 1.0$&$97.8$ $\pm 2.1$&$98.1$ $\pm 1.7$&$98.5$ $\pm 1.4$&$97.9$ $\pm 1.3$& $79.8$ $\pm 2.3$\\
TV-CovNet\_152\_sc\_mean& \bm{$99.8$} $\pm 0.3$&\bm{$99.6$} $\pm 0.5$&\bm{$99.9$} $\pm 0.4$&\bm{$99.7$} $\pm 0.4$&{$99.7$} $\pm 0.3$&\bm{$84.4$} $\pm 1.6$\\
\hline
\end{tabular}
\end{table*}
\begin{table*}[!ht]
\caption{Performance on COVID-19 Diagnosis With Chest X-ray Imaging (two classes).}
\label{tab:coviddetails} \centering
\begin{tabular}{p{87 pt}|p{\smallcolwidth pt}|p{\smallcolwidth pt}|p{\smallcolwidth pt}|p{\smallcolwidth pt}|p{\smallcolwidth pt}|p{\smallcolwidth pt}|p{\smallcolwidth pt}|p{\smallcolwidth pt}|p{\smallcolwidth pt}|p{\smallcolwidth pt}|p{\smallcolwidth pt}|p{\smallcolwidth pt}|p{\smallcolwidth pt}|p{\smallcolwidth pt}|p{\smallcolwidth pt}|p{20 pt}|p{25 pt}}
Split & 1 & 2 & 3 & 4& 5 & 6& 7 & 8& 9 & 10& 11 & 12& 13 & 14& 15 & Average&p-value\\
\hline
\multicolumn{18}{c}{two classes}\\
\hline
{ResNet-50}&98.6&97.7&99.6&98.1&$\bm{99.5}$&97.8&97.2&98.6&98.6&98.6&99.1&99.5&98.2&99.5&100&98.7&-\\
\hline
TV-CovNet\_50\_sc\_mean&$\bm{100}$&$\bm{99.5}$&$\bm{100}$&$\bm{99.5}$&${99.1}$&$\bm{100}$&$\bm{99.5}$&$\bm{100}$&$\bm{100}$&$\bm{100}$&$\bm{99.5}$&$\bm{100}$&$\bm{100}$&$\bm{99.5}$&$\bm{100}$&$\bm{99.8}$&5.3x10$^{-5}$\\
\hline
\hline
{ResNet-101}&99.1&99.1&95.7&97.6&98.6&$\bm{99.6}$&97.7&98.2&99.5&99.0&100&99.1&98.6&99.0&99.5&98.7&-\\
\hline
TV-CovNet\_101\_sc\_mean&$\bm{100}$&$\bm{100}$&$\bm{99.1}$&$\bm{100}$&$\bm{100}$&${98.7}$&$\bm{99.1}$&$\bm{100}$&$\bm{100}$&$\bm{99.5}$&$\bm{99.5}$&$\bm{99.5}$&$\bm{100}$&$\bm{99.5}$&$\bm{99.5}$&$\bm{99.6}$&3.5x10$^{-3}$\\
\hline
\hline
{ResNet-152}&97.7&100&96.1&98.6&97.2&97.4&96.8&99.1&100&98.1&99.1&99.5&98.6&99.5&98.1&98.4&-\\
\hline
TV-CovNet\_152\_sc\_mean&$\bm{100}$&$\bm{100}$&$\bm{99.6}$&$\bm{100}$&$\bm{99.5}$&$\bm{99.1}$&$\bm{99.5}$&$\bm{100}$&$\bm{100}$&$\bm{100}$&$\bm{99.5}$&$\bm{100}$&$\bm{100}$&$\bm{100}$&$\bm{99.5}$&$\bm{99.8}$&1.3x10$^{-4}$\\
\hline
\multicolumn{18}{c}{three classes}\\
\hline
{ResNet-50}&80.0&78.1&81.4&77.7&80.0&78.2&75.5&81.1&$\bm{84.2}$&80.1&80.2&79.9&77.1&80.1&76.7&79.4&-\\
\hline
TV-CovNet\_50\_sc\_mean&$\bm{84.9}$&$\bm{84.1}$&$\bm{84.8}$&$\bm{84.7}$&$\bm{86.8}$&$\bm{84.4}$&$\bm{80.9}$&$\bm{82.4}$&${82.1}$&$\bm{86.2}$&$\bm{81.8}$&$\bm{83.6}$&$\bm{85.8}$&$\bm{83.7}$&$\bm{79.6}$&$\bm{83.7}$&4.4x10$^{-6}$\\
\hline
\hline
{ResNet-101}&80.8&77.3&81.9&75.7&$\bm{83.0}$&78.0&74.7&76.5&81.5&79.3&78.1&77.5&78.9&81.2&76.9&78.8&-\\
\hline
TV-CovNet\_101\_sc\_mean&$\bm{81.4}$&$\bm{82.4}$&$\bm{85.9}$&$\bm{82.3}$&${81.1}$&$\bm{82.0}$&$\bm{83.2}$&$\bm{82.7}$&$\bm{83.2}$&$\bm{87.6}$&$\bm{83.6}$&$\bm{81.2}$&$\bm{85.0}$&$\bm{83.4}$&$\bm{79.6}$&$\bm{83.0}$&1.9x10$^{-5}$\\
\hline
\hline
{ResNet-152}&78.9&78.9&79.1&77.4&82.5&80.1&76.3&76.5&81.0&77.3&79.9&80.7&75.8&79.5&77.5&78.8&-\\
\hline
TV-CovNet\_152\_sc\_mean&$\bm{83.2}$&$\bm{84.1}$&$\bm{84.3}$&$\bm{85.6}$&$\bm{85.8}$&$\bm{85.1}$&$\bm{81.6}$&$\bm{84.6}$&$\bm{87.5}$&$\bm{83.1}$&$\bm{83.6}$&$\bm{83.9}$&$\bm{83.7}$&$\bm{87.3}$&$\bm{82.3}$&$\bm{84.4}$&2.4x10$^{-9}$\\
\hline
\end{tabular}
\end{table*}

\subsection{Diagnostic Results}
The above section verifies the assumption of the proposed TV-CovNet, and this section will evaluate its effectiveness to integrate the left lung, right lung and overall views for joint diagnosis. For simplicity and clear comparison, an identical backbone network architecture is applied to the three steams of TV-CovNet. The network architecture is set as ResNet-50, ResNet-101 and ResNet-152, respectively, to verify its generality with respect to different network depths. Both binary classification, i.e., between Normal  and COVID-19, and three class classification among Normal,  COVID-19 and Other will be performed. The results on binary case will be evaluated by five metrics with true positive (TP), true negative (TN), false positive (FP), and false negative (FN) values, including 
\begin{itemize}
\item Accuracy = (TN + TP)/(TN + TP + FN + FP)
\item Precision = TP/(TP + FP)
\item Recall = TP/(TP + FN)
\item Specificity = TN/(TN + FP) 
\item F1 score = 2(Precision $\times$ Recall)/(Precision + Recall)
\end{itemize}
In the three class case, accuracy is used as the evaluation metric. As aforementioned, the random partition of training/test sets are repeated 15 times for each task to obtain stable statistics. Table~\ref{tab:covidsummary} reports the results in four portions, with the first portion containing the competing methods, including COVID-Net~\cite{wang2020covid} and ResNet-18 based LocalPatch~\cite{oh2020deep} method. The rest three portions report the performance of methods with ResNet-50, ResNet-101 and ResNet-152~\cite{he2016deep}, respectively. In each of these three portions, we firstly show the performance of the baseline method, i.e., a single stream Resnet with the overall view as input. It is followed by the results of LocalPatch~\cite{oh2020deep} method with ResNet. Then we report the performance of the proposed TV-CovNet with five fusion methods, where ``fea'' denotes feature level fusion and ``sc'' refers to score level fusion introduced in Section~\ref{sectionfusion}, while ``cat'', ``max'' and ``mean'' indicate feature concatenation, max pooling and mean pooling, respectively. 

As reported in the first portion, the competing method COVID-Net~\cite{wang2020covid} obtains an accuracy of 78.0\% in the three class classification case. Its is not applied to the two class case since the CNN in that method is intentionally designed for the three class classification task. The ResNet-18 based LocalPatch~\cite{oh2020deep} method improves the performance to 80.5\% in the three class classification case, achieving an improvement of 2.5 percentage points.

In the second portion, as seen in the binary classification case, ResNet-50~\cite{he2016deep} already achieves very promising accuracy of 98.7\%. LocalPatch~\cite{oh2020deep} with ResNet-50 improves the performance of ResNet-50 in all the five metrics, verifying the effectiveness of generating patches for major voting in \cite{oh2020deep}. When the proposed TV-CovNet is applied with ResNet-50 architecture as backbone networks, among the feature fusion methods, TV-CovNet\_fea\_cat and TV-CovNet\_fea\_max obtain comparable accuracy to the baseline ResNet-50~\cite{he2016deep}, but no improvement is observed. In contrast, TV-CovNet\_fea\_mean improves ResNet-50 in all the five metrics. This demonstrates the effectiveness of integrating triple-view information for joint diagnosis. When score fusion methods are applied, TV-CovNet\_sc\_max is comparable to ResNet-50~\cite{he2016deep} while TV-CovNet\_50\_sc\_mean outperforms ResNet-50~\cite{he2016deep}, becoming comparable  to {ResNet-50 + LocalPatch}~\cite{oh2020deep}. 

In the more challenging three class classification task, a similar trend can be observed. Specifically, TV-CovNet\_fea\_cat, TV-CovNet\_fea\_max and TV-CovNet\_sc\_max are comparable to ResNet-50 while TV-CovNet\_fea\_mean and TV-CovNet\_50\_sc\_mean achieve considerable improvement over ResNet-50. Especially, TV-CovNet\_50\_sc\_mean outperforms all the competing methods, achieving an improvement of 4.3 percentage points over ResNet-50 and 2.1 percentage points over the state-of-the-art method LocalPatch~\cite{oh2020deep}.

Similar results can be observed with ResNet-101 and ResNet-152 architectures as backbone networks, as shown in the bottom two portions of Table~\ref{tab:covidsummary}. Especially, the proposed mean score pooling (denoted as sc\_mean) method consistently obtains the best performance among the competing methods.

In order to provide more details on the results above, the accuracies of the baseline ResNet and TV-CovNet with mean score pooling in 15 splits are shown in Table~\ref{tab:coviddetails}, including both of the binary classification case and the three class classification case. As seen, regardless of ResNet-50, ResNet-101 or ResNet-152 is used, the proposed TV-CovNet\_sc\_mean outperforms ResNet in most splits and the improvement is statistically evident, as verified by the small p-value ($<0.05$) of student's t-test. The corresponding confusion matrices averaged over 15 splits are shown in  Table~\ref{tab:covidconfusionmatrix2c}, including both of the binary classification case and the three class classification case. As can be seen that all the diagonal entries of the confusion matrices in the right column obtained by the proposed method are enlarged while all the off-diagonal entries are reduced in comparison with the results of baseline ResNet in the left column. This observation is consistent in both binary and three class cases with any of ResNet-50, -101 or -152.

In summary, as verified in our experiments above, fusion at the score layer performs better than fusion at the feature level in the proposed TV-CovNet. Regarding the fusion methods, mean pooling performs better than feature concatenation or max pooling methods. TV-CovNet with mean pooling at the score level consistently outperforms the competing methods and demonstrates the state-of-the-art performance.

\newcommand\itemstwo{2}   %Number of classes
\arrayrulecolor{white} %Table line colors
\newcommand\items{3}   %Number of classes

\begin{table}[!ht]
\caption{Confusion Matrices of Different methods for COVID-19 Diagnosis With Chest X-ray Imaging (N - {Normal}, C - COVID-19, O - Other).}
\label{tab:covidconfusionmatrix2c} \centering

\begin{tabular}{c|c}
\multicolumn{2}{c}{Two class case}\\
~\\
\noindent\begin{tabular}{cc*{\itemstwo}{|E}|}
\multicolumn{1}{c}{} &\multicolumn{1}{c}{} &\multicolumn{\itemstwo}{c}{Predicted} \\ \hhline{~*\itemstwo{|-}|}
\multicolumn{1}{c}{} & 
\multicolumn{1}{c}{} & 
\multicolumn{1}{c}{\rot{N}} & 
\multicolumn{1}{c}{\rot{C}} \\ \hhline{~*\itemstwo{|-}|}
\multirow{\itemstwo}{*}{\rotatebox{90}{Actual}} 
&N& 128.5   & 1.5     \\ \hhline{~*\itemstwo{|-}|}
&C& 2   & 85    \\ \hhline{~*\itemstwo{|-}|}
%TV-CovNet\_50\_sc\_mean&TV-CovNet\_101\_sc\_mean&TV-CovNet\_152\_sc\_mean\\
%ResNet-50&ResNet-101&ResNet-152\\
\end{tabular}&
\noindent\begin{tabular}{cc*{\itemstwo}{|E}|}
\multicolumn{1}{c}{} &\multicolumn{1}{c}{} &\multicolumn{\itemstwo}{c}{Predicted} \\ \hhline{~*\itemstwo{|-}|}
\multicolumn{1}{c}{} & 
\multicolumn{1}{c}{} & 
\multicolumn{1}{c}{\rot{N}} & 
\multicolumn{1}{c}{\rot{C}} \\ \hhline{~*\itemstwo{|-}|}
\multirow{\itemstwo}{*}{\rotatebox{90}{}} 
&N& 129.6   & 0.4     \\ \hhline{~*\itemstwo{|-}|}
&C& 0.07   & 86.93    \\ \hhline{~*\itemstwo{|-}|}
%TV-CovNet\_50\_sc\_mean&TV-CovNet\_101\_sc\_mean&TV-CovNet\_152\_sc\_mean\\
%ResNet-50&ResNet-101&ResNet-152\\
\end{tabular}\\
~\\
ResNet-50&TV-CovNet\_50\_sc\_mean\\
\noindent\begin{tabular}{cc*{\itemstwo}{|E}|}
\multicolumn{1}{c}{} &\multicolumn{1}{c}{} &\multicolumn{\itemstwo}{c}{} \\ \hhline{~*\itemstwo{|-}|}
\multicolumn{1}{c}{} & 
\multicolumn{1}{c}{} & 
\multicolumn{1}{c}{\rot{N}} & 
\multicolumn{1}{c}{\rot{C}} \\ \hhline{~*\itemstwo{|-}|}
\multirow{\itemstwo}{*}{\rotatebox{90}{Actual}} 
&N& 128.3   & 1.7     \\ \hhline{~*\itemstwo{|-}|}
&C& 1.2   & 85.8    \\ \hhline{~*\itemstwo{|-}|}
%TV-CovNet\_50\_sc\_mean&TV-CovNet\_101\_sc\_mean&TV-CovNet\_152\_sc\_mean\\
%ResNet-50&ResNet-101&ResNet-152\\
\end{tabular}&
\noindent\begin{tabular}{cc*{\itemstwo}{|E}|}
\multicolumn{1}{c}{} &\multicolumn{1}{c}{} &\multicolumn{\itemstwo}{c}{} \\ \hhline{~*\itemstwo{|-}|}
\multicolumn{1}{c}{} & 
\multicolumn{1}{c}{} & 
\multicolumn{1}{c}{\rot{N}} & 
\multicolumn{1}{c}{\rot{C}} \\ \hhline{~*\itemstwo{|-}|}
\multirow{\itemstwo}{*}{\rotatebox{90}{}} 
&N& 129.7   & 0.3    \\ \hhline{~*\itemstwo{|-}|}
&C& 0.5   &  86.5   \\ \hhline{~*\itemstwo{|-}|}
%TV-CovNet\_50\_sc\_mean&TV-CovNet\_101\_sc\_mean&TV-CovNet\_152\_sc\_mean\\
%ResNet-50&ResNet-101&ResNet-152\\
\end{tabular}\\
~\\
ResNet-101&TV-CovNet\_101\_sc\_mean\\
\noindent\begin{tabular}{cc*{\itemstwo}{|E}|}
\multicolumn{1}{c}{} &\multicolumn{1}{c}{} &\multicolumn{\itemstwo}{c}{} \\ \hhline{~*\itemstwo{|-}|}
\multicolumn{1}{c}{} & 
\multicolumn{1}{c}{} & 
\multicolumn{1}{c}{\rot{N}} & 
\multicolumn{1}{c}{\rot{C}} \\ \hhline{~*\itemstwo{|-}|}
\multirow{\itemstwo}{*}{\rotatebox{90}{Actual}} 
&N& 128.5   & 1.5     \\ \hhline{~*\itemstwo{|-}|}
&C& 2   & 85    \\ \hhline{~*\itemstwo{|-}|}
%TV-CovNet\_50\_sc\_mean&TV-CovNet\_101\_sc\_mean&TV-CovNet\_152\_sc\_mean\\
%ResNet-50&ResNet-101&ResNet-152\\
\end{tabular}&
\noindent\begin{tabular}{cc*{\itemstwo}{|E}|}
\multicolumn{1}{c}{} &\multicolumn{1}{c}{} &\multicolumn{\itemstwo}{c}{} \\ \hhline{~*\itemstwo{|-}|}
\multicolumn{1}{c}{} & 
\multicolumn{1}{c}{} & 
\multicolumn{1}{c}{\rot{N}} & 
\multicolumn{1}{c}{\rot{C}} \\ \hhline{~*\itemstwo{|-}|}
\multirow{\itemstwo}{*}{\rotatebox{90}{}} 
&N& 129.7   & 0.3     \\ \hhline{~*\itemstwo{|-}|}
&C& 0.1   & 86.9    \\ \hhline{~*\itemstwo{|-}|}
%TV-CovNet\_50\_sc\_mean&TV-CovNet\_101\_sc\_mean&TV-CovNet\_152\_sc\_mean\\
%ResNet-50&ResNet-101&ResNet-152\\
\end{tabular}\\
~\\
ResNet-152&TV-CovNet\_152\_sc\_mean\\
~\\
\multicolumn{2}{c}{Three class case}\\
~\\
\noindent\begin{tabular}{cc*{\items}{|E}|}
\multicolumn{1}{c}{} &\multicolumn{1}{c}{} &\multicolumn{\items}{c}{Predicted} \\ \hhline{~*\items{|-}|}
\multicolumn{1}{c}{} & 
\multicolumn{1}{c}{} & 
\multicolumn{1}{c}{\rot{N}} & 
\multicolumn{1}{c}{\rot{C}} & 
\multicolumn{1}{c}{\rot{O}} \\ \hhline{~*\items{|-}|}
\multirow{\items}{*}{\rotatebox{90}{Actual}} 
&N& 103.0   & 1.3  & 25.7   \\ \hhline{~*\items{|-}|}
&C& 0.9   & 77.9  & 8.3   \\ \hhline{~*\items{|-}|}
&O& 27.0   & 13.5   & 113.9   \\ \hhline{~*\items{|-}|}
\end{tabular}&
\noindent\begin{tabular}{cc*{\items}{|E}|}
\multicolumn{1}{c}{} &\multicolumn{1}{c}{} &\multicolumn{\items}{c}{Predicted} \\ \hhline{~*\items{|-}|}
\multicolumn{1}{c}{} & 
\multicolumn{1}{c}{} & 
\multicolumn{1}{c}{\rot{N}} & 
\multicolumn{1}{c}{\rot{C}} & 
\multicolumn{1}{c}{\rot{O}} \\ \hhline{~*\items{|-}|}
\multirow{\items}{*}{\rotatebox{90}{}} 
&N& 113.5   & 0.6  & 15.9   \\ \hhline{~*\items{|-}|}
&C& 0.4   & 79.4  & 7.3   \\ \hhline{~*\items{|-}|}
&O& 23.3   & 13.1   & 118.1   \\ \hhline{~*\items{|-}|}
\end{tabular}\\
~\\
ResNet-50&TV-CovNet\_50\_sc\_mean\\
\noindent\begin{tabular}{cc*{\items}{|E}|}
\multicolumn{1}{c}{} &\multicolumn{1}{c}{} &\multicolumn{\items}{c}{} \\ \hhline{~*\items{|-}|}
\multicolumn{1}{c}{} & 
\multicolumn{1}{c}{} & 
\multicolumn{1}{c}{\rot{N}} & 
\multicolumn{1}{c}{\rot{C}} & 
\multicolumn{1}{c}{\rot{O}} \\ \hhline{~*\items{|-}|}
\multirow{\items}{*}{\rotatebox{90}{Actual}} 
&N& 102.9   & 2.9  & 24.2   \\ \hhline{~*\items{|-}|}
&C& 0.7   & 78.5  & 7.1   \\ \hhline{~*\items{|-}|}
&O& 28.3   & 15.5   & 110.3   \\ \hhline{~*\items{|-}|}
\end{tabular}&
\noindent\begin{tabular}{cc*{\items}{|E}|}
\multicolumn{1}{c}{} &\multicolumn{1}{c}{} &\multicolumn{\items}{c}{} \\ \hhline{~*\items{|-}|}
\multicolumn{1}{c}{} & 
\multicolumn{1}{c}{} & 
\multicolumn{1}{c}{\rot{N}} & 
\multicolumn{1}{c}{\rot{C}} & 
\multicolumn{1}{c}{\rot{O}} \\ \hhline{~*\items{|-}|}
\multirow{\items}{*}{\rotatebox{90}{}} 
&N& 110.1   & 1  & 18.9   \\ \hhline{~*\items{|-}|}
&C& 0.7   & 79.5  & 6.7   \\ \hhline{~*\items{|-}|}
&O& 23.0   & 13.6   & 118.5   \\ \hhline{~*\items{|-}|}
\end{tabular}\\
~\\
ResNet-101&TV-CovNet\_101\_sc\_mean\\
\noindent\begin{tabular}{cc*{\items}{|E}|}
\multicolumn{1}{c}{} &\multicolumn{1}{c}{} &\multicolumn{\items}{c}{} \\ \hhline{~*\items{|-}|}
\multicolumn{1}{c}{} & 
\multicolumn{1}{c}{} & 
\multicolumn{1}{c}{\rot{N}} & 
\multicolumn{1}{c}{\rot{C}} & 
\multicolumn{1}{c}{\rot{O}} \\ \hhline{~*\items{|-}|}
\multirow{\items}{*}{\rotatebox{90}{Actual}} 
&N& 105.6   & 1.8  & 22.6   \\ \hhline{~*\items{|-}|}
&C& 1.1   & 80.0  & 6.7   \\ \hhline{~*\items{|-}|}
&O&31.2   & 15.5   & 107.1   \\ \hhline{~*\items{|-}|}
\end{tabular}&
\noindent\begin{tabular}{cc*{\items}{|E}|}
\multicolumn{1}{c}{} &\multicolumn{1}{c}{} &\multicolumn{\items}{c}{} \\ \hhline{~*\items{|-}|}
\multicolumn{1}{c}{} & 
\multicolumn{1}{c}{} & 
\multicolumn{1}{c}{\rot{N}} & 
\multicolumn{1}{c}{\rot{C}} & 
\multicolumn{1}{c}{\rot{O}} \\ \hhline{~*\items{|-}|}
\multirow{\items}{*}{\rotatebox{90}{}} 
&N& 113.9   & 0.5  & 15.6   \\ \hhline{~*\items{|-}|}
&C& 0.4   & 80.1  & 6.5   \\ \hhline{~*\items{|-}|}
&O& 22.0   & 13.0   & 119.5   \\ \hhline{~*\items{|-}|}
\end{tabular}\\
~\\
ResNet-152&TV-CovNet\_152\_sc\_mean\\
\end{tabular}
\end{table}
\arrayrulecolor{black} %Table line colors

\section{Discussion}
\subsection{Comparison with Ensemble Methods}
The above section has verified the effectiveness of the proposed TV-CovNet, which integrates the visual features from the left lung, right lung and overall views. A natural question arises that, if the key objective is to integrate the complementary information from the three views for joint diagnosis, would ensemble methods also serve this purpose? To answer this question, we train three classifiers separately with the the cropped left lung, right lung and overall views, respectively, and ensemble the output scores of these three classifiers during test phase by applying max or mean functions. The results are reported in Table \ref{tab:ensemble}. As seen, the ensemble method with max, denoted by ensemble\_max, or mean, denoted by ensemble\_mean, consistently improves the performance of baseline ResNet in both of the two and three class classification cases no matter ResNet-50, ResNet-101 or ResNet-152 is used. However, the ensemble methods do not perform as well as the proposed TV-CovNet\_sc\_mean method since the later could better integrate the complementary information from the three views in an end-to-end learning manner. In contrast, the ensemble methods treat the three views independently and do not allow interaction between them during the training stage.

\begin{table}[!ht]
\caption{Comparison between ensemble and the proposed methods.}
\label{tab:ensemble} \centering
\begin{tabular}{l|cc}
					 & $2$ classes & $3$ classes \\
Methods in comparison & Accuracy & Accuracy \\
\toprule
{ResNet-50}&$98.7$ $\pm 0.8$& $79.4$ $\pm 2.2$\\
\hline
ResNet-50\_ensemble\_max& $99.0$ $\pm 0.6$& $81.2$ $\pm 1.9$\\
ResNet-50\_ensemble\_mean& $99.0$ $\pm 0.8$& $82.8$ $\pm 1.6$\\
TV-CovNet\_50\_sc\_mean (proposed)& \bm{$99.8$} $\pm 0.3$& \bm{$83.7$} $\pm 2.0$\\
\hline
\hline
{ResNet-101}&$98.7$ $\pm 1.1$& $78.8$ $\pm 2.5$\\
\hline
ResNet-101\_ensemble\_max& $ 99.1$ $\pm 0.8$& $81.1$ $\pm 2.2$\\
ResNet-101\_ensemble\_mean& $99.2$ $\pm 0.5$& $82.4$ $\pm 2.0$\\
TV-CovNet\_101\_sc\_mean (proposed)& \bm{$99.6$} $\pm 0.4$& \bm{$83.0$} $\pm 2.0$\\
\hline
\hline
{ResNet-152}&$98.4$ $\pm 1.2$& $78.8$ $\pm 1.9$\\
\hline
ResNet-152\_ensemble\_max& $98.8$ $\pm 0.7$& $81.3$ $\pm 1.8$\\
ResNet-152\_ensemble\_mean& $98.9$ $\pm 0.9$& $82.9$ $\pm 1.9$\\
TV-CovNet\_152\_sc\_mean (proposed)& \bm{$99.8$} $\pm 0.3$&\bm{$84.4$} $\pm 1.6$\\
\hline
\end{tabular}
\end{table}

\subsection{Comparison with Double-view Networks}
This section will investigate is the overall view still required in TV-CovNet considering that the left lung and right lung views already contain all the visual information in the two lungs? To this end, we remove the overall view stream from the proposed TV-CovNet and train double-view networks, denoted as DV-CovNet, with the left lung and right lung only. Its performance is compared with TV-CovNet. Since the mean score pooling has consistently obtained the best performance among the five fusion methods in all the experiments above, only it is applied in this section. As reported in Table~\ref{tab:coviddouble}, DV-CovNet indeed improves the baseline, ResNet, but it is outperformed by the proposed TV-CovNet counterpart. This demonstrates that the overall view is indeed contributive in TV-CovNet. This is probably because, although the left and right lung views contain all the visual contents of the two lungs, the overall view enables learning certain co-occurrence features and forming a hierarchical representation of an CXR image, which could benefit final diagnosis.

\begin{table}[!ht]
\caption{Comparison between double-view and the proposed methods.}
\label{tab:coviddouble} \centering
\begin{tabular}{l|cc}
					 & $2$ classes & $3$ classes \\
Methods in comparison & Accuracy & Accuracy \\
\toprule
{ResNet-50}&$98.7$ $\pm 0.8$& $79.4$ $\pm 2.2$\\
\hline
DV-CovNet\_50\_sc\_mean& $99.2$ $\pm 0.8$& $82.2$ $\pm 2.8$\\
TV-CovNet\_50\_sc\_mean (proposed)& \bm{$99.8$} $\pm 0.3$& \bm{$83.7$} $\pm 2.0$\\
\hline
\hline
{ResNet-101}&$98.7$ $\pm 1.1$& $78.8$ $\pm 2.5$\\
\hline
DV-CovNet\_101\_sc\_mean& $99.4$ $\pm 0.6$& $81.8$ $\pm 2.4$\\
TV-CovNet\_101\_sc\_mean (proposed)& \bm{$99.6$} $\pm 0.4$& \bm{$83.0$} $\pm 2.0$\\
\hline
\hline
{ResNet-152}&$98.4$ $\pm 1.2$& $78.8$ $\pm 1.9$\\
\hline
DV-CovNet\_152\_sc\_mean& $99.1$ $\pm 0.6$& $82.0$ $\pm 1.6$\\
TV-CovNet\_152\_sc\_mean (proposed)& \bm{$99.8$} $\pm 0.3$&\bm{$84.4$} $\pm 1.6$\\
\hline
\end{tabular}
\end{table}

\subsection{If pretrained networks help?}
In the experiments above, the ResNet parameters are initialized with the pretrained networks on ImageNet~\cite{imagenet_cvpr09}, however, the images from ImageNet are natural images rather than medical images. This section will verify does initialization with pretrained parameters benefit the diagnosis accuracy given the huge domain gap between the images from ImageNet and  those from the CXR dataset constructed in this paper. Table~\ref{tab:covidnopretrained} reports the performance of ResNet and TV-CovNet with or without pretrained parameter initialization. As seen, initialization with pretrained parameters always outperforms random initialization in both cases of the ResNet and the proposed TV-CovNet with any of ResNet-50, -101 or -152.

\begin{table}[!ht]
\caption{Comparison between non-pretrained and pretrained initialization.}
\label{tab:covidnopretrained} \centering
\begin{tabular}{l|cc}
					 &$2$ classes & $3$ classes \\
Methods in comparison & Accuracy& Accuracy \\
\toprule
{ResNet-50 (non-pretrained)}& $98.3$ $\pm 0.8$& $77.9$ $\pm 2.3$\\
{ResNet-50 (pretrained)}& \bm{$98.7$} $\pm 0.8$& \bm{$79.4$} $\pm 2.2$\\
\hline
TV-CovNet\_50\_sc\_mean (non-pretrained)& {$98.5$} $\pm 0.8$& {$79.9$} $\pm 1.6$\\
TV-CovNet\_50\_sc\_mean (pretrained)& \bm{$99.8$} $\pm 0.3$& \bm{$83.7$} $\pm 2.0$\\
\hline
\hline
{ResNet-101 (non-pretrained)}& $98.5$ $\pm 0.8$& $78.4$ $\pm 1.7$\\
{ResNet-101 (pretrained)}& \bm{$98.7$} $\pm 1.1$& \bm{$78.8$} $\pm 2.5$\\
\hline
TV-CovNet\_101\_sc\_mean (non-pretrained)& {$98.7$} $\pm 0.8$& {$79.8$} $\pm 1.7$\\
TV-CovNet\_101\_sc\_mean (pretrained)& \bm{$99.6$} $\pm 0.4$& \bm{$83.0$} $\pm 2.0$\\
\hline
\hline
{ResNet-152 (non-pretrained)}& $98.0$ $\pm 0.9$& $78.3$ $\pm 1.7$\\
{ResNet-152 (pretrained)}& \bm{$98.4$} $\pm 1.2$& \bm{$78.8$} $\pm 1.9$\\
\hline
TV-CovNet\_152\_sc\_mean (non-pretrained)& {$98.6$} $\pm 0.6$& {$80.1$} $\pm 1.9$\\
TV-CovNet\_152\_sc\_mean (pretrained)& \bm{$99.8$} $\pm 0.3$& \bm{$84.4$} $\pm 1.6$\\
\hline
\end{tabular}
\end{table}

\subsection{Weighted fusion via adaptive learning}
It has been verified above that, among the five fusion methods in \ref{sectionfusion}, the mean score pooling method leads to the best performance in TV-CovNet. In that method, the three views are treated equally important. This section aims to study will weighted fusion be able to further improve the performance. To this end, three positive scalars ${w}_{L}$, ${w}_{O}$, ${w}_{R}$ are assigned to the corresponding scores of three streams $\bm{s}_{L}$, $\bm{s}_{O}$, $\bm{s}_{R}$, respectively, as weight of the steam. These weights are adaptively optimized during the training stage. The weighted score mean, i.e., $\frac{1}{{w}_{L} + {w}_{O} + {w}_{R}}\Sigma_{i \in \{L, O, R\}}{w_{i}\bm{s}_{i}}$, is used as the final prediction score. As reported in Table~\ref{tab:covidadaptive}, the weighted mean performs worse than TV-CovNet\_sc\_mean. This is probably due to the limited  training data scale, and the adaptively learned weights may lead to over-fitting in this case.

\begin{table}[!ht]
\caption{Comparison between dynamically-learned weighted combination  methods and the proposed methods.}
\label{tab:covidadaptive} \centering
\begin{tabular}{l|cc}
					 &$2$ classes & $3$ classes \\
Methods in comparison & Accuracy& Accuracy \\
\toprule
{ResNet-50}& $98.7$ $\pm 0.8$& $79.4$ $\pm 2.2$\\
\hline
TV-CovNet\_50\_sc\_mean (weighted)& {$99.2$} $\pm 0.7$& {$82.6$} $\pm 1.9$\\
TV-CovNet\_50\_sc\_mean (proposed)& \bm{$99.8$} $\pm 0.3$& \bm{$83.7$} $\pm 2.0$\\
\hline
\hline
{ResNet-101}& $98.7$ $\pm 1.1$& $79.4$ $\pm 2.2$\\
\hline
TV-CovNet\_101\_sc\_mean (weighted)& {$99.5$} $\pm 0.6$& {$82.7$} $\pm 2.2$\\
TV-CovNet\_101\_sc\_mean (proposed)& \bm{$99.6$} $\pm 0.4$& \bm{$83.0$} $\pm 2.0$\\
\hline
\hline
{ResNet-152}& $98.4$ $\pm 1.2$& $79.4$ $\pm 2.2$\\
\hline
TV-CovNet\_152\_sc\_mean (weighted)& {$99.3$} $\pm 0.5$& {$82.1$} $\pm 2.1$\\
TV-CovNet\_152\_sc\_mean (proposed)& \bm{$99.8$} $\pm 0.3$& \bm{$84.4$} $\pm 1.6$\\
\hline
\end{tabular}
\end{table}

\subsection{Integration with other methods}
The proposed TV-CovNet is an flexible framework, which can be easily integrated with various network architectures or training strategies. This section shows an integration of the proposed method with {LocalPatch}~\cite{oh2020deep} as an example. Specifically, we generate random region crops from each of the three views and use these crops as input to train TV-CovNet by following {LocalPatch}~\cite{oh2020deep}. During the test stage, we also generate 100 crops for each view and apply major voting to obtain the final label as in \cite{oh2020deep}. As seen in Table~\ref{tab:combinewithpatch}, the integrated method, TV-CovNet\_101\_sc\_mean + LocalPatch, further improves the diagnosis accuracy in the three class classification task with any of ResNet-50, -101 or -152. The two class case is not reported since the existing methods already achieved very high accuracy and the challenging three class case could better demonstrate the comparison.

\begin{table}[!ht]
\caption{Combination of patch based and the proposed methods.}
\label{tab:combinewithpatch} \centering
\begin{tabular}{l|c}
					 & $3$ classes \\
Methods in comparison & Accuracy \\
\toprule
{ResNet-50}~\cite{he2016deep}& $79.4$ $\pm 2.2$\\
{ResNet-50 + LocalPatch}~\cite{oh2020deep}& $81.6$ $\pm 1.8$\\
TV-CovNet\_50\_sc\_mean& {$83.7$} $\pm 2.0$\\
TV-CovNet\_50\_sc\_mean + LocalPatch& \bm{$86.4$} $\pm 1.6$\\
\hline
{ResNet-101}~\cite{he2016deep}& $78.8$ $\pm 2.5$\\
{ResNet-101 + LocalPatch}~\cite{oh2020deep}& $81.1$ $\pm 1.7$\\
TV-CovNet\_101\_sc\_mean& {$83.0$} $\pm 2.0$\\
TV-CovNet\_101\_sc\_mean + LocalPatch& \bm{$85.4$} $\pm 2.1$\\
\hline
{ResNet-152}~\cite{he2016deep}& $78.8$ $\pm 1.9$\\
{ResNet-152 + LocalPatch}~\cite{oh2020deep}& $81.4$ $\pm 2.1$\\
TV-CovNet\_152\_sc\_mean& {$84.4$} $\pm 1.6$\\
TV-CovNet\_152\_sc\_mean + LocalPatch& \bm{$85.5$} $\pm 1.7$\\
\hline
\end{tabular}
\end{table}

\subsection{Study on Training Data Scale}
\begin{figure}[!htb]
\begin{center}
{\includegraphics[width = 80 mm]{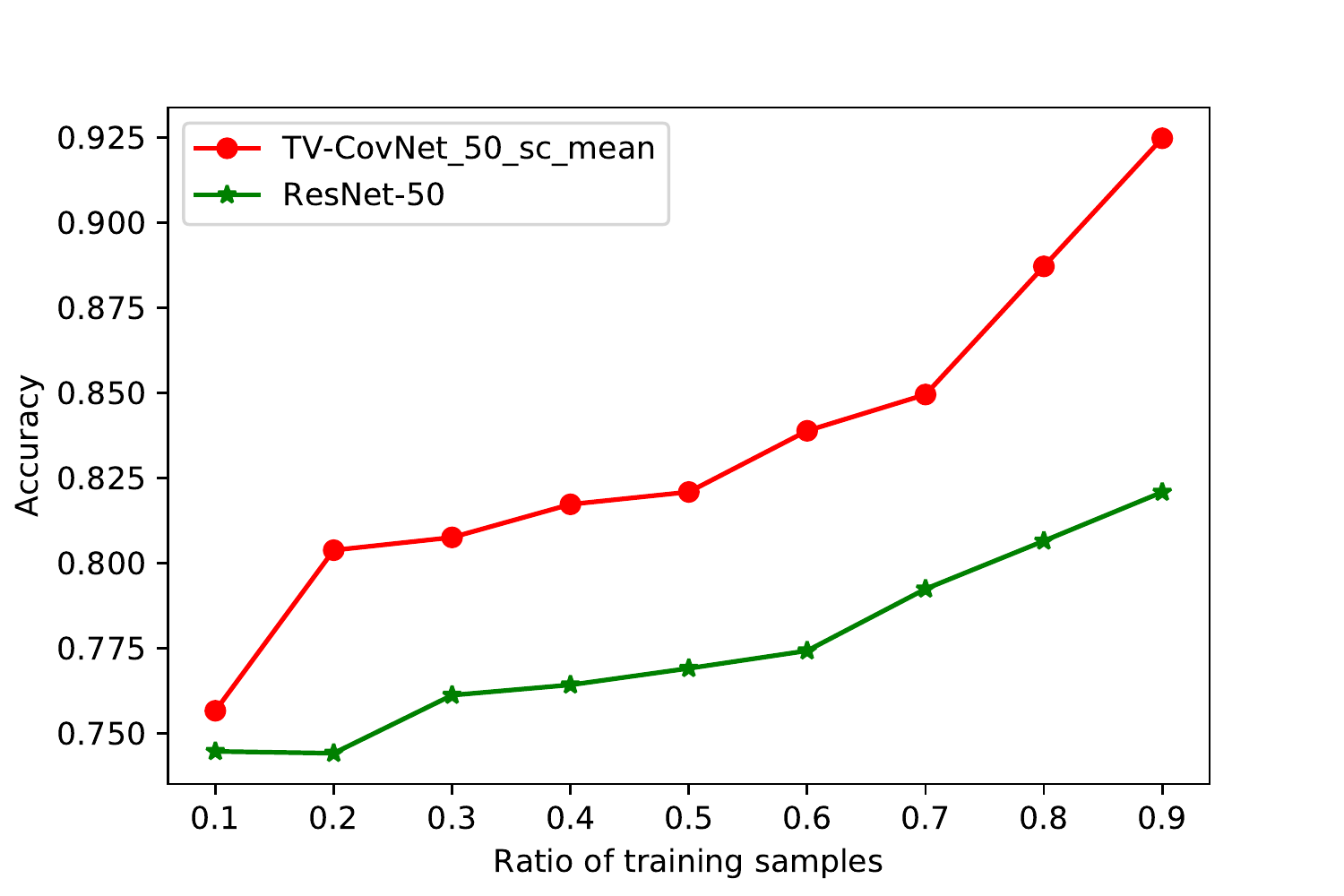}}
\end{center}
%\vspace{-6 mm}
\caption{The comparison between the proposed TV-CovNet\_50\_sc\_mean and ResNet-50 with different ratios of training samples.}
\label{fig:ratios}
\end{figure}

This section studies the performance trend of the baseline ResNet and proposed TV-CovNet with respect to different scales of training data in the three class classification case. As shown in Fig~\ref{fig:ratios}, the x-axis indicates the ratios of samples assigned to training set and y-axis shows the corresponding performance of ResNet-50 and the proposed TV-CovNet\_50\_sc\_mean. As seen, with increasing ratios, although both methods achieve higher accuracy, TV-CovNet\_50\_sc\_mean admits faster accuracy increase and its improvement over ResNet-50 becomes larger. This study indicates the promising potential learning capacity of TV-CovNet.

\section{Conclusion and future work}
In order to better extract informative visual features from the two lungs in CXR images for COVID-19 diagnosis, we proposed a triple-view network structure. The proposed structure respects the anatomical structure of human lungs and is well aligned with clinician's diagnosis practice with CXR images. The advantages and effectiveness of the proposed structure are experimentally verified in both binary classification task between normal and COVID-19 cases and three class classification task among normal, COVID-19 and non-COVID-19 pneumonia. All the results consistently show that the proposed structure obtains the state-of-the-art performance. Various properties of the proposed method are discussed, including its comparison with ensemble methods, its flexibility to be extended and promising modeling capacity with increasing training data scale etc. These discussions present more insights on why the proposed method performs well and may inspire future explorations in this line.

Several open issues are worth exploring along this line of research. Firstly, the proposed networks can be extended to other scan modalities, e.g., CT, or a combination of multiple scans. Secondly, the backbone networks in different streams are not necessarily the same. Specially designed networks for each stream may admit more adaptive feature extraction and achieve better diagnosis accuracy. Last but not least, collecting larger scale of training data for the proposed method is also critical for further evaluation and performance improvement.

\ifCLASSOPTIONcaptionsoff
  \newpage
\fi

\bibliographystyle{IEEEtran}
\bibliography{covid19}

\end{document}